# Search for neutrinoless double beta decay with NEMO 3 experiment

Zornitza Daraktchieva on behalf of NEMO collaboration

*Department of Physics and Astronomy, University College London, Gower Street, London, WC1E 6BT, UK*

**Abstract**
NEMO 3 experiment is designed to search for neutrinoless double beta decay ($\beta\beta 0\nu$). It is located in the Modane Underground Laboratory (LSM) and has been taking data since February 2003. The half-lives of two neutrino double beta decay ($\beta\beta 2\nu$) have been measured for seven isotopes. No evidence of neutrinoless beta decay has been found. The limits on both the half-lives of the $\beta\beta 0\nu$ and the corresponding Majorana effective masses are derived.

## 1. Introduction

The neutrinoless double beta decay ($\beta\beta 0\nu$) is a golden channel to probe the Majorana nature of the neutrino. The main objective of the NEMO 3 experiment (Neutrino Ettore Majorana Observatory) is to search for $\beta\beta 0\nu$ decay and to measure accurately the two neutrino double beta decay ($\beta\beta 2\nu$) for several double beta isotopes. $\beta\beta 2\nu$ decay is allowed in the Standard model: it is a process in which two neutrons are transformed into two protons with emission of two electrons and two neutrinos. By contrast, $\beta\beta 0\nu$ decay is forbidden in the Standard Model: it is a process with emission of only two electrons in the final state. The main idea behind the NEMO 3 experiment is to measure accurately the energy and tracks of the electrons in the final state by employing both calorimetric and tracking detection techniques.

## 2. NEMO 3 detector

The NEMO 3 cylindrical detector [1], divided in 20 sectors, houses enriched isotopes in the form of thin source foils. Seventeen sectors enclose $^{100}$Mo (6914 g), $^{82}$Se(932 g), $^{116}$Cd(405 g), $^{130}$Te(454 g), $^{150}$Nd(34 g), $^{96}$Zr(9.4 g) and $^{48}$Ca(7 g). The three remaining sectors contain pure Cu and natural Te for external background measurements. The tracking and calorimeter techniques have been employed for detecting the electrons in the final state, i.e., the experimental signature of $\beta\beta 0\nu$. The source foils are fixed vertically between two gaseous tracking detectors, made of 6180 open octagonal drift cells operating in Geiger mode, providing a three dimensional reconstruction of the charged particle tracks. The energy and time of flight measurements of particles are determined by a calorimeter made of 1940 plastic scintillators coupled to 3" and 5" PMTs. Electrons and positrons are identified by a 25 Gauss magnetic field parallel to the foil axis.

A double beta decay candidate is a two-electron event that answers the following criteria: two long tracks coming from the same vertex in the source foils with negative curvature, two scintillator hits associated with tracks and the time of flight corresponding to the case of two electrons emitted at the same time from the same vertex. The background in the two neutrino double beta decay sample has been directly measured using different channels of analysis.

### 3. NEMO 3 results

The two neutrino double beta decay of seven isotopes is measured accurately with NEMO 3 experiment. Here are presented the recent results for $^{150}$Nd [2]. The data collected between February 2003 and December 2006 correspond to 924.7 days of data. 2789 two-electron double beta candidates have been selected with 7.2 % efficiency, with 746 background events expected. Figure 1 (left) shows the energy distribution of the sum of two electrons for $^{150}$Nd. The data are in a good agreement with the simulated distribution of two neutrino double beta decay signal and the background. The measured half-life of $^{150}$Nd is found to be
$T_{1/2}^{\beta\beta 2\nu} = \left[ 9.11^{+0.025}_{-0.022}(stat.) \pm 0.63(syst.) \right] \times 10^{18} \, y$. Table 1 shows the half-lives of the other six isotopes.

| Isotope | $Q_{\beta\beta}$ (MeV) | $T_{1/2} \times 10^{19}$ (y) | Signal/Background |
|---|---|---|---|
| $^{100}$Mo | 3.034 | 0.711 ± 0.002 (stat) ± 0.054 (syst) [3] | 40 |
| $^{82}$Se | 2.295 | 9.6 ± 0.3 (stat) ± 1.0 (syst) [3] | 4 |
| $^{130}$Te | 2.529 | 76 ± 15 (stat) ± 8 (syst) [4] | 0.25 |
| $^{116}$Cd | 2.805 | 2.8 ± 0.1 (stat) ± 0.3 (syst) [4] | 7.5 |
| $^{96}$Zr | 3.350 | 2.3 ± 0.2 (stat) ± 0.3 (syst) | 1.0 |
| $^{48}$Ca | 4.772 | $4.4^{+0.5}_{-0.4}$(stat) ± 0.4 (syst) | 6.8 |

Table 1. Measured half-lives of ββ2ν decay by NEMO 3 experiment.

A neutrinoless double beta decay signal would correspond to an excess of two electron type events in the distribution of the energy sum of the electrons around the energy of the transition $Q_{\beta\beta}$(3.367 MeV for $^{150}$Nd). No excess of events in the distribution of the energy sum has been observed during 924.7 days. The energy distribution for $E_{sum} > 2.5$ MeV is given in Figure 1 (right) for data compared to the background. MC simulation of a neutrinoless double beta decay signal is also shown.

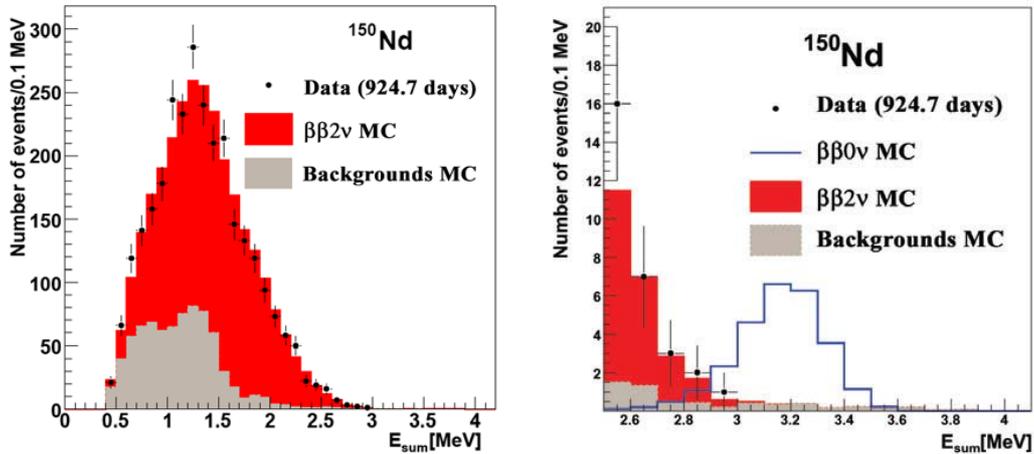

Fig.1 Spectra of energy sum of two electrons (left) and energy sum of two electrons above 2.5 MeV in *ββ0ν* energy window (right).The Monte Carlo simulations of neutrinoless double beta decay signal is given as well (right).

A limit on the half-life of *ββ0ν* is found to be $T_{1/2}^{\beta\beta 0\nu} > 1.8 \times 10^{22}\, y$ (90% C.L.), given that the CLs method [5] is employed. This limit corresponds to the limit on the effective Majorana neutrino mass $m_\nu < 1.5 - 2.5$ eV calculated with nuclear matrix elements (NME) from QPRA without taking into account the nucleus deformation [6,7]. Note that in the case of nuclear deformation, derived from laboratory moments [8], the suppression of the NME for $^{150}$Nd has been estimated to be a factor 2.7. This increases the upper limit to $m_\nu < 4.0 - 6.3$ eV which is consistent with the limit derived using the NME of a pseudo SU(3) model [9] and the PHFB model [10], which also include the effect of nuclear deformation. The limits on the half-lives of *ββ0ν* and corresponding Majorana masses of the other NEMO 3 isotopes are given in Table 2.

| Isotope | $T_{1/2}$ (y) (90 % C.L.) | $m_\nu$ (eV) | NME references |
|---|---|---|---|
| $^{100}$Mo | > 5.8 x $10^{23}$ | < 0.6÷1.3 [3] | [6,7,11,12,13]] |
| $^{82}$Se | > 2.1 x $10^{23}$ | < 1.2÷2.2 [3] | [6,7,11,12,13] |
| $^{96}$Zr | > 8.6 x $10^{21}$ | < 7.4÷20.1 | [6,7,11,12,13] |
| $^{48}$Ca | > 1.3 x $10^{22}$ | < 29.7 | [14] |

Table 2. Limits on the half-lives of *ββ0ν* decay and corresponding Majorana masses